# Multiscale Entropy Analysis of the Portevin-Le Chatelier Effect in an Al-2.5%Mg Alloy

A. Sarkar, P. Barat and P. Mukherjee

Variable Energy Cyclotron Centre

1/AF Bidhan Nagar, Kolkata 700064, India

**Abstract:**
*The complexity of the Portevin-Le Chatelier effect in Al-2.5%Mg polycrystalline samples subjected to uniaxial tensile tests is quantified. Multiscale entropy analysis is carried out on the stress time series data observed during jerky flow to quantify the complexity of the distinct spatiotemporal dynamical regimes. It is shown that for the static type C band, the entropy is very low for all the scales compared to the hopping type B and the propagating type A bands. The results are interpreted considering the time and length scales relevant to the effect.*

Portevin-Le Chatelier (PLC) effect, observed in many dilute alloys of technological importance, is one of the widely studied metallurgical phenomena. It is a striking example of the complexity of spatiotemporal dynamics resulting from the collective behavior of dislocations. In uniaxial loading with constant imposed strain rate, the effect manifests itself as a series of repeated stress drops (serrations) in the stress-time or strain curve. Each stress drop is associated with the nucleation of a localized band of plastic deformation [1, 2]. The band propagates along the sample under certain conditions. The microscopic origin of the PLC effect is the dynamic strain aging (DSA) of the material due to the interaction between the mobile dislocations and the diffusing solute atoms [3,4]. Mobile dislocations which are the carrier of plastic strain, move jerkily between the obstacles provided by the other defects. Solute atoms diffuse in the stress field generated by the mobile dislocations. In a certain range of strain rates and temperatures, the diffusion time of the solute atoms is of the order of the waiting time of the mobile dislocations temporarily arrested at the obstacles. In this ranges of the state variables, solute atoms diffuse to the dislocations and age them. At the macroscopic scale, this DSA leads to a negative strain rate sensitivity of the flow stress and makes the plastic deformation nonuniform [3-5]. In polycrystals three generic types of serrations namely type A, B and C are distinguished [6, 7]. With increasing strain rate type C, B and A serrations are observed chronologically. These three types of serrations are associated with distinct features of the deformation bands. At low strain rates type C serrations are identified with randomly nucleated static band with large characteristic load drops [1, 2]. These bands are associated with weak spatial coupling [7]. At medium strain rates relatively smaller serrations (type B) are observed. The bands formed in this case exhibit an oscillatory or intermittent propagation (hopping) with marginal spatial coupling. At high strain rates bands propagate continuously and type A serrations with small stress drops are observed [1, 2]. The type A bands are strongly correlated.

Beyond its importance in metallurgy, the PLC effect is a paradigm for a general class of nonlinear complex systems with intermittent bursts [8]. The succession of plastic instabilities shares both physical and statistical properties with many other systems exhibiting loading-unloading cycles e.g. earthquake. PLC effect is regulated by interacting mechanisms that operate across multiple spatial and temporal scales. The output variable (stress) of the effect exhibits complex fluctuations which contains information about the underlying dynamics.

Many statistical and dynamical studies have been carried out on the PLC effect over the last two decades [9-14]. These studies have provided a considerable understanding of the underlying dynamics of the PLC effect. Analysis revealed two types of dynamical regimes in the PLC effect. At medium strain rate (type B) chaotic regime has been demonstrated [12, 13], which is associated with the bell-shaped distribution of the stress drops. For high strain rate (type A) the dynamics is identified as self organized criticality (SOC) [13] with the stress drops following a power law distribution. The crossover of these two mechanisms has also been a topic of intense research for the past few years [10, 13, 15].

To the best of our knowledge no effort has been made to quantify the complexity of the PLC effect observed at different strain rates. Quantifying the complexity of the time series from a physical process may be of considerable interest due to its potential application in evaluating a dynamical model of the system. In this letter we present a quantitative study of the measure of complexity of the PLC effect at the different strain rates. Entropy based algorithm are often used to quantify the regularity of a time series [16, 17]. Increase in entropy corresponds to the increase in the degree of disorder and for a completely random system it is maximum. Traditional algorithms are single-scale based [16, 17]. However, time series derived from the complex systems are likely to present structure on multiple temporal scales. In contrast, time series derived from a simpler system are likely to present structures on just a single scale. For these reasons the traditional single scale based algorithms often yield misleading quantifications of the complexity of a system.

Recently Costa et al. [19] introduced a new method, Multiscale entropy (MSE) analysis for measuring the complexity of finite length time series. This method measures complexity taking into account the multiple time scales. This computational tool can be quite effectively used to quantify the complexity of a natural time series. The first multiple scale measurement of the complexity was proposed by Zhang [20]. Zhang's method was based on the Shannon entropy which requires a large number of almost noise free data. On the contrary, the MSE method uses Sample Entropy (SampEn) to quantify the regularity of finite length time series. SampEn is largely independent of the time series length when the total number of data points is larger than approximately 750 [21]. Thus MSE proved to be quite useful in analyzing the finite length time series over the Zhang's method. Recently MSE has been successfully applied to quantify the complexity of many Physiologic and Biological signals [19, 22, 23]. Here, we take the initiative to apply this novel method to quantify the complexity of a metallurgical phenomenon.

Al-Mg alloys containing a nominal percentage of Mg exhibit the PLC effect at room temperature for a wide range of strain rates [6]. We have carried out tensile tests on flat specimens prepared from polycrystalline Al-2.5%Mg alloy. Specimens with gauge length, width, and thickness of 25, 5 and 2.3 mm, respectively, were tested in a servo controlled INSTRON (model 4482) machine. All the tests were carried out at room temperature (300K) and at a sampling rate of 20Hz. We have carried out the tests at three different strain rates: $8.0 \times 10^{-6}$ sec$^{-1}$, $2.0 \times 10^{-4}$ sec$^{-1}$ and $1.9 \times 10^{-3}$ sec$^{-1}$. These strain rates were chosen in such a way that we could observe only type C, type B and type A serrations respectively in the stress time data [6]. The stress time data obtained from these experiments show an increasing drift due to the strain hardening effect. To remove the effect of strain hardening we have subtracted the drift by the method of polynomial fitting. The investigations presented below were carried out on the resulting time series. The segments of the stress time curves for the three strain rates are shown in Fig. 1.

The MSE method is based on the evaluation of SampEn on the multiple scales. The prescription of the MSE analysis is: given a one-dimensional discrete time series, $\{x_1,......,x_i,....,x_N\}$, construct the consecutive coarse-grained time series, $\{y^{(\tau)}\}$, determined by the scale factor, $\tau$, according to the equation:

$$y_j^\tau = 1/\tau \sum_{i=(j-1)\tau+1}^{j\tau} x_i \qquad (1)$$

where $\tau$ represents the scale factor and $1 \le j \le N/\tau$. The length of each coarse-grained time series is $N/\tau$. For scale one, the coarse-grained time series is simply the original time series. Next we calculate the SampEn for each scale using the following method. Let $\{X_i\} = \{x_1,......,x_i,......,x_N\}$ be a time series of length N. $u_m(i) = \{x_i, x_{i+1},......,x_{i+m-1}\}, 1 \le i \le N-m$ be vectors of length $m$. Let $n_{im}(r)$ represent the number of vectors $u_m(j)$ within distance $r$ of $u_m(i)$, where $j$ ranges from 1 to (N-m) and $j \ne i$ to exclude the self matches. $C_i^m(r) = n_{im}(r)/(N-m-1)$ is the probability that any $u_m(j)$ is within $r$ of $u_m(i)$. We then define

$$U^m(r) = 1/(N-m) \sum_{i=1}^{N-m} \ln C_i^m(r) \qquad (2)$$

The parameter Sample Entropy (SampEn) [21] is defined as

$$SampEn(m,r) = \lim_{N \to \infty} \left\{ -\ln \frac{U^{m+1}(r)}{U^m(r)} \right\} \qquad (3)$$

For finite length N the SampEn is estimated by the statistics

$$SampEn(m,r,N) = -\ln \frac{U^{m+1}(r)}{U^m(r)} \qquad (4)$$

Advantage of SampEn is that it is less dependent on time series length and is relatively consistent over broad range of possible r, m and N values. We have calculated SampEn for all the studied data sets with the parameters m=2 and r= 0.15×SD (SD is the standard deviation of the original time series).

Costa et al. had tested the MSE method on simulated white and 1/f noises [19]. They have shown that for the scale one, the value of entropy is higher for the white noise time series in comparison to the 1/f noise, as shown in Fig. 2. This may apparently lead to the conclusion that the inherent complexity is more in the white noise in comparison to the 1/f noise. However, the application of the MSE method shows that the value of the entropy for the 1/f noise remains almost invariant for all the scales while the value of entropy for the white noise time series monotonically decreases and for scales > 5, it becomes smaller than the corresponding values for the 1/f noise. This result explains the fact that the 1/f noise contains complex structures across multiple scales in contrast to the white noise. With a view to understand the complexity of a chaotic process we have generated chaotic data from the logistic map $x_{n+1}= x_n (1-x_n)$ with a=3.9 and applied the MSE method. The entropy measure for the chaotic time series increases on small scales and then gradually decreases indicating the reduction of complexity on the larger scales, as shown in Fig. 2.

We next apply the MSE method to the analysis of the stress fluctuations observed in the PLC effect. The results of the MSE analysis of the stress time series for the three studied strain rates are shown in Fig. 3. It is seen that for the lowest strain rate, where only the type C bands are observed, the SampEn is very low for all the scales. The entropy measure slightly increases with the scale. This signifies the simplicity of the type C band dynamics compared to the other two bands. For the medium strain rate i.e. for the type B serrations, the entropy measure markedly increases in the small scales and then gradually decreases and gets saturated. The entropy measure for the high strain rate stress time series (type A serrations) decreases on the small scale and then gradually increases. For scale one, the entropy assigned to the stress time series for the type A serrations is higher than that of the type B and C stress time series. In contrast for scales > 3 the time series of type B serrations are assigned to the highest entropy values. For the largest scale, the entropy measures for the type B and the type A serrations become almost equal. Moreover, for the type B serrations the variation of the SampEn with the scale is similar to that of the chaotic data [Fig. 2]. This result corroborates with the previous finding of. the presence of deterministic chaos in the type B band dynamics [12, 13]. The high values of entropy for all the scales for the type A stress time series data may be a signature of the SOC as observed earlier [13].

A possible approach for understanding the observed complexity measure is to consider the properties of the time and length scales relevant to the PLC effect. At low strain rates, the reloading time between two successive drops, $t_l$, is very large compared to the plastic relaxation time $t_r$ i.e. $t_l \gg t_r$. The internal stresses are fully relaxed in this case and consequently the type C bands are spatially uncorrelated. The absence of spatial correlation makes the dynamics of the type C band simple and it may be easier to build a model based on the repeated pinning and unpinning of dislocations in the field of solute atoms. When the driving strain rate is increased, the reloading time scale $t_l$ decreases and approaches the relaxation time scale $t_r$. Internal stresses are not totally relaxed during the reloading periods. Hence, new bands are formed nearby the previous ones leading to the hopping character of the associated type B bands. This spatial coupling and the competition between the two time scales increases the complexity of the type B band dynamics. At high strain rates the ratio $t_l/t_r$ decreases, which hinders the plastic relaxation. Very little plastic relaxation occurs between the stress drops. Thus, the stress felt by the dislocations always remain close to the threshold for unpinning from the solute atmosphere. New bands are formed in the field of the unrelaxed internal stresses and perennial plastic events overlap resulting in a hierarchy of length scales. This leads to both SOC dynamics and type A propagating bands [24]. Due to this hierarchy the entropy measure for the high strain rate comes out to be high for all the scales. The high complexity of the type B and A band dynamics makes the modeling of PLC effect very difficult.

In recent years there have been a great enthusiasm to model the PLC effect. Some good models [1, 2, 7], which reproduce some of the features of the PLC effect very well have also been emerged. However, there are constant efforts to provide better models. Knowledge of this complexity measure should be taken care of while modeling the PLC effect in different strain rate region.

In conclusion we have carried out the Multi Scale Entropy analysis on the Portevin-Le Chatelier effect. The entropy measure is very low for the low strain rate i.e. for type C serrations, signifying the simplicity of the type C band dynamics. For medium strain rate the entropy increases markedly in the small scales and then gradually decreases and gets saturated. For the high strain rate the entropy decreases in small scales and then gradually increases with the scale and finally at the largest scale it becomes almost equal to the entropy measure of the medium strain rate. The study clearly establishes the fact that the PLC dynamics of the type A and B bands is the most complex one.

We thank Dr. M. Costa of Harvard Medical School for helpful discussions and for providing the 1/f noise data.

**Figure captions**

Fig. 1. Segments of the stress time curves at three strain rates : (a) $8.0\times10^{-6}$ sec$^{-1}$, (b) $2.0\times10^{-4}$ sec$^{-1}$ and (c) $1.9\times10^{-3}$ sec$^{-1}$.

Fig. 2. MSE analysis of white noise, 1/f noise and logistic map chaotic data. On the y axis, the value of the entropy (SampEn) for the coarse-grained time series is plotted. The scale factor specifies the number of data points averaged to obtain each element of the coarse-grained time series.

Fig. 3. MSE analysis of the stress time series for the strain rates: $8.0\times10^{-6}$ sec$^{-1}$ (low), $2.0\times10^{-4}$ sec$^{-1}$ (medium), and $1.9\times10^{-3}$ sec$^{-1}$ (high).

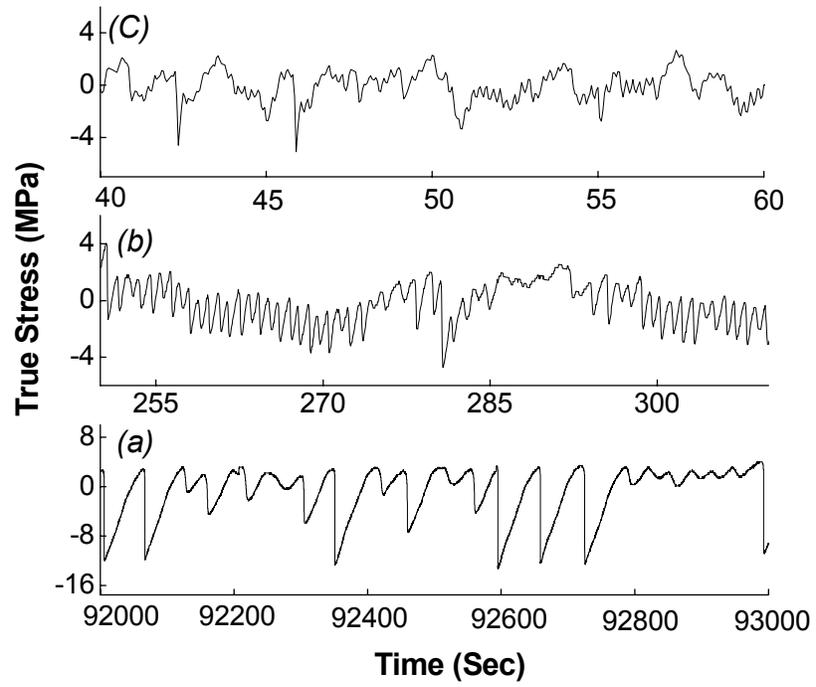

Fig. 1

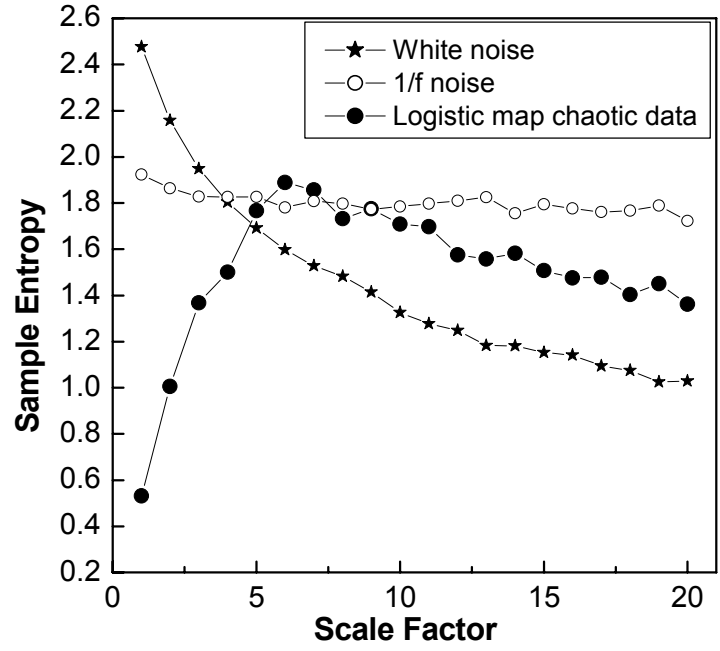

Fig. 2

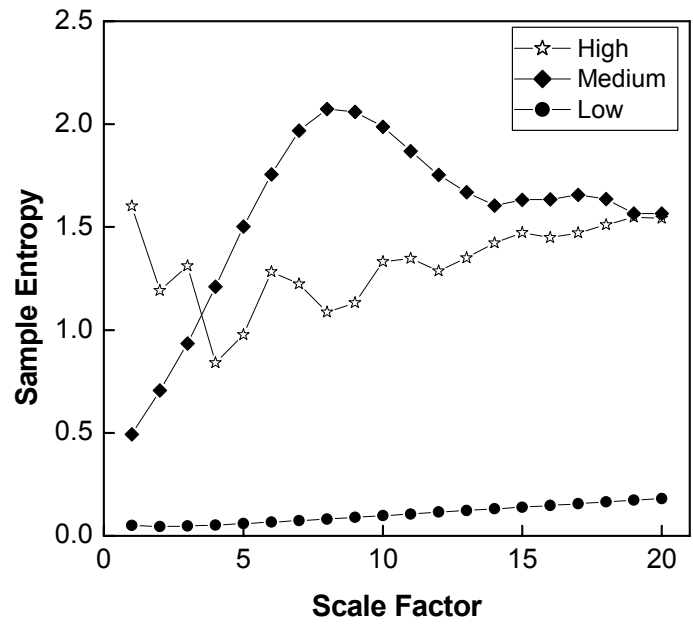

Fig. 3